# Large topological Hall effect near room temperature in noncollinear ferromagnet LaMn$_2$Ge$_2$ single crystal

Gaoshang Gong,[1,2,‡] Longmeng Xu,[1,†] Yuming Bai[1], Yongqiang Wang[2], Songliu Yuan[1], Yong Liu[3], Zhaoming Tian[1*]

[1] School of Physics and Wuhan National High Magnetic Field Center, Huazhong University of Science and Technology, Wuhan 430074, P. R. China

[2] School of Physics and Electronic Engineering, Zhengzhou University of Light Industry, Zhengzhou 450002, P. R. China

[3] School of Physics, Wuhan University, Wuhan 430072, P. R. China

**ABSTRACT**

Non-trivial spin structures in itinerant magnets can give rise to topological Hall effect (THE) due to the interacting local magnetic moments and conductive electrons. While, in series of materials, THE has mostly been observed at low temperatures far below room temperature (RT) limiting its potential applications. Here, we report the anisotropic anomalous Hall effect (AHE) near RT in LaMn$_2$Ge$_2$, a noncollinear ferromagnetic (FM) with Curie temperature $T_C$~325 K. Large topological Hall resistivity of ~1.0 $\mu\Omega \cdot$cm in broad temperature range (190 K <T< 300 K) is realized as field ($H$) parallel to the *ab*-plane ($H // ab$) and current along *c* axis ($I // c$), in contrast to the conventional AHE for $H // c$ and $I // ab$. The emergence of THE is attributed to the spin chirality of noncoplanar spin configurations stabilized by thermal fluctuation during spin flop process. Moreover, the constructed temperature-field ($H$-$T$) phase diagrams based on the isothermal topological Hall resistivity reveal a field-induced transition from the noncoplanar spin configuration to polarized ferromagnetic states. Our experimental realization of large THE near RT highlights LaMn$_2$Ge$_2$ as a promising system for functional applications in novel spintronic devices.

---

[‡] These authors contributed equally to this work.
\* Corresponding author should to be address: tianzhaoming@hust.edu.cn

## I. INTRODUCTION

The interplay between charge and spin degrees of freedom in itinerant magnets can generate various kinds of intriguing electromagnetic phenomena. As one typical example, the anomalous Hall effect (AHE) characterized by a transverse voltage generated by longitudinal charge current even at zero field in ferromagnetic (FM) metals [1-3], has attracted considerable attention from both the fundamental physics of magnetotransport and spintronic applications. As for its origin, two qualitatively different mechanisms are widely accepted, extrinsic impurity scattering processes and intrinsic contribution from band structure [3-5]. The extrinsic mechanisms involve the skew-scattering and side-jump scattering ones [6,7], which produce the AHE due to the asymmetric scattering of conduction electrons. The intrinsic Karplus-Luttinger mechanism is related to the spin-orbit interaction (SOI) and perturbation by the electric field initially proposed by Karplus and Luttinger [8], recent theories have reinterpreted it by invoking the Berry-phase concepts [9,10]. Compared to the AHE in conventional metallic ferromagnets, another unconventional Hall signal termed topological Hall effect (THE) can also arise in materials with nontrivial spin structures [11-14]. In such systems, the conduction electrons can acquire a nonzero Berry phase associated with finite scalar spin chirality $\chi_{ijk}=S_i \cdot (S_j \times S_k)$ when passing through the localized spin moments ($S_i$, $S_j$, and $S_k$) [11,12], which acts as an emergent magnetic field responsible for the THE. In this mechanism, the magnitude of THE is related to the $\chi_{ijk}$ not $M$. Therefore, the total Hall resistivity ($\rho_{xy}$) in itinerant magnets should consist three contributions expressed by $\rho_{xy} = R_0\mu_0 H + R_S M + \rho_{xy}^T$, (1) where the first, second and third terms represent the normal, anomalous and topological Hall resistivity, respectively. $R_0$ and $R_S$ are the normal and anomalous Hall coefficients, respectively.

In experiment, THE has been widely reported in different systems hosting topologically nontrivial spin textures such as in magnetic skyrmions [13-15], double-exchanged ferromagnets [16,17], frustrated magnets [12,18-21] as well as artificial magnetic heterostructures [22,23]. Among them, frustrated magnets as ideal platform where the spins usually form noncollinear or noncoplanar configurations with nonzero $\chi_{ijk}$ are of particularly attractive to explore THE, as reported in pyrochlore lattice $Nd_2Mo_2O_7$ and $Pr_2Ir_2O_7$ [11,12], triangular lattice $PdCrO_2$ and $Gd_2PdSi_3$ [24,25], Kagome lattice $Mn_3Sn$ and $Fe_3Sn_2$ [26-28], etc. On the other side, the summation of $\chi_{ijk}$ over the whole lattice sites in frustrated magnets can also be macroscopically canceled out due to the lack of chiral magnetic symmetry, such as in ideal 120º spin structure [18,29], thereby THE is not a common feature among frustrated magnets. Moreover, limited by the low transition temperature, THE in most materials appears at temperatures far below room temperature (RT) hindering its practical applications, experimental realizations of large THE at RT in new materials are highly appealing for its direct applications in spintronic devices.

$LaMn_2Ge_2$ belongs to a family of layered intermetallic compounds $RM_2X_2$ crystallized in

ThCr$_2$Si$_2$-type structure with centrosymmetric tetragonal space group *I*4/mmm (R: rare-earth element, M: 3d or 4d element and X: Si or Ge) [30], where magnetic Mn atoms are located on square-lattice sheets stacked parallelly along the *c* axis [see Figs. 1(a) and 1(b)]. Previous neutron scattering studies revealed the Mn moments formed the conical spin structure of below Curie temperature $T_C$~325 K [31], which couple ferromagnetically along easy *c* axis and antiferromagnetically within the *ab*-plane. Considering the noncollinear magnetic ordering well above RT, LaMn$_2$Ge$_2$ provides a candidate to identify RT THE, motivating our present study.

Here, we investigate the anisotropic AHE up to RT in LaMn$_2$Ge$_2$ single crystals based on the systematical measurements of magnetic and electrical transport for field (*H*) along different directions. Compared to the observation of conventional AHE as *H* along *c* axis (*H* // *c*) and current (*I*) within the *ab*-plane (*I* // *ab*), large topological hall resistivity of ~1.0 μΩ·cm extended up to RT is observed under *H* // *ab* and *I* // *c* axis. The emergence of THE is attributed to the scalar spin chirality related to the formation of noncoplanar spin configurations stabilized by thermal fluctuation during spin flop process.

## II. EXPERIMENTAL DETAILS

Single crystals of LaMn$_2$Ge$_2$ were grown from the Indium flux. High-purity elements La(99.9%) and Mn(99.95%), Ge(99.99%) and In(99.99%) from Alfa Aesar in the molar ratio La:Mn:Ge:In=1:2:2:20 were put into an alumina crucible and sealed inside an evacuated ampoule. Then, it was heated at 1100°C for 12 h, and cooled at a rate of 4°C/h to 700°C. At this temperature, the ampoule was taken out from the furnace rapidly and the Indium flux was decanted by a centrifuge, plate like crystals were obtained in a typical dimension of 0.5×3×3 mm$^3$. The structure was characterized by X-ray diffraction (XRD) collected with Cu $K_\alpha$ radiation at room temperature by using a diffractometer (Rigaku-TTR3), which confirmed the purity of samples.

Magnetic and electrical transport measurement were carried out using the superconducting quantum interference device magnetometer (SQUID, Quantum Design) and commercial Physical Property Measurement System (PPMS, Quantum Design), respectively. A standard four-probe method was used to perform the longitudinal and transverse electrical transport measurements with *H* always perpendicular to *I* direction. Both the out-of-plane ($\rho_{zz}$, $\rho_{zx}$) and in-plane ($\rho_{xx}$, $\rho_{xy}$) electrical transports were measured with *I* // *c* axis and *I* // *ab*-plane, respectively. To eliminate the influence of contact electrode misalignments, Hall resistivity was measured for both field directions and symmetrized by $\rho_H(\mu_0 H) = [\rho_H(+\mu_0 H) - \rho_H(-\mu_0 H)]/2$. All electrical transport measurements were repeated using crystals from the same batch with similar residual resistivity ratio (RRR), guaranteeing the reliability of experimental results.

## III. RESULTS AND DISCUSSIONS

LaMn$_2$Ge$_2$ has a tetragonal structure with lattice constant $a=b=4.19$ Å and $c=10.95$ Å, where La, Mn and Ge atoms occupy the Wyckoff positions 2a (0,0,0) 4d (0,0.5,0.25) and 4(e) (0,0,0.38), respectively. As depicted in Fig. 1(a), the La, Mn and Ge atoms are located on separate layers alternating along *c* axis in a sequence: -Mn-Ge-La-Ge-Mn-, where magnetic Mn layers are well separated by the nonmagnetic La and Ge layers [30]. In this structure, the MnGe$_4$ tetrahedra are connected in an edge-sharing fashion with Mn atom at the center of each tetrahedron. Projection of LaMn$_2$Ge$_2$ on the (001) plane [shown in Fig. 1(b)], one can see the magnetic Mn atoms lay out on square lattice within the *ab* plane. The nearest Mn-Mn intralayer distance given by $a/\sqrt{2} = 2.97$ Å is much smaller than that of interlayer separation $c/2 = 5.48$ Å. For the crystals, the *c*-axis is normal to the plane of plate.

Temperature (*T*) dependence of magnetic susceptibilities χ(T) for LaMn$_2$Ge$_2$ were measured under *H* = 0.1 T along *a* and *c* axis ($\chi_a$ and $\chi_c$), respectively. For both directions, zero-field-cooled (ZFC) and field-cooled (FC) magnetization curves nearly overlap with slight differences at low temperatures. As shown in Fig. 1(c), a rapid upturn of $\chi_c$ is observed as *T* close to 325 K, characterizing the paramagnetic-ferromagnetic (PM-FM) transition at $T_C$ ~325 K. Below $T_C$, $\chi_c$ is nearly ten times larger than $\chi_a$, denoting the easy magnetization along *c* axis. The zero-field in-plane $\rho_{xx}$ (*I // ab*) and out-of-plane resistivity $\rho_{zz}$ (*I // c*) are also presented in Fig. 1(d), both curves exhibit metallic behaviors in all temperature regimes with large residual resistivity ratio (RRR) ρ(375 K)/ρ(10 K)~24. A clear anomaly at ~325 K corresponds to the $T_C$ determined from χ(T) curves. The ratio of $\rho_{zz}(T)/\rho_{xx}(T)$~0.6-0.65 is almost temperature independent, reveals the in-plane and out-of-plane transport share the same scattering mechanism. The weak anisotropy implies three-dimensional (3D) electrical transport property rather than two-dimensional (2D) ones.

Figures. 2(a) and (b) exhibit the isothermal magnetization (*M*) curves at different temperatures for *H // c* and *H // a* axis, respectively. The magnetizations are plotted versus effective magnetic field ($\mu_0 H_{eff}$) defined by $\mu_0 H_{eff} = \mu_0(H - N_d M)$, where $N_d$ is the demagnetization factor for samples in rectangular shape calculated by a method in Ref. 32. For both directions with *T<<T$_C$*, the magnetizations increase linearly at low fields and become saturation as field above the saturation field ($\mu_0 H_s$). At *T* = 10 K, different saturation fields $\mu_0 H_s^c = 0.32$ T (*H // c*) and $\mu_0 H_s^a = 2.1$ T (*H // a*) reveal that LaMn$_2$Ge$_2$ has uniaxial magnetic anisotropy with easy axis along *c* direction. The saturated magnetization $M_S$~3.12 $\mu_B$/f.u. is consistent with the previous reported value [33]. It is worthwhile to mention that we also measure the isothermal $M(\mu_0 H_{eff})$ for *H* along different directions within the *ab*-plane, which show almost isotropic behaviors.

LaMn$_2$Ge$_2$ displays strong anisotropic magnetotransport behaviors as *H* oriented along the *c* axis and within the *ab*-plane. In Figs. 3(a) and 3(b), we firstly present the in-plane longitudinal magnetoresistance (MR) and Hall resistivity $\rho_{xy}(\mu_0 H_{eff})$, respectively. During the measurements, the current is along *a* axis and field along *c* axis. Negative MRs $\{MR = [\rho(\mu_0 H) - \rho(0T)]/\rho(0T)\}$ are observed at high temperatures with maximum near $T_C$ (see Fig. 3(a) and Fig. S1 in Supplementary Material [34]), due to the suppression of spin-related scattering as usually observed in magnetic systems. As decreased temperature, MR changes sign to positive at low temperatures (*T* < 60 K), reflecting the dominant contribution of MR from the Lorenz force induced by *H* on the carrier motion. For field dependent Hall resistivity $\rho_{xy}(\mu_0 H_{eff})$, it has a linear field dependence at $T > T_C$. Below $T_C$, $\rho_{xy}(\mu_0 H_{eff})$ increases dramatically at low fields, then becomes slowly at high fields with linear field dependence. As comparisons, the shape of $\rho_{xy}(\mu_0 H_{eff})$ resembles to the $M(\mu_0 H_{eff})$ curve, suggest the dominated conventional anomalous Hall effect (AHE) as observed in FM metals. In this respect, Hall resistivity can be described by $\rho_{xy} = \rho_{xy}^N + \rho_{xy}^A = R_0 \mu_0 H + R_S \mu_0 M$, where $\rho_{xy}^N$, $\rho_{xy}^A$, $R_0$ and $R_S$ represent the normal and anomalous Hall resistivity, normal and anomalous Hall coefficients, respectively. The value of $R_0$ and $\rho_{xy}^A$ can be determined from the linear fit of $\rho_{xy}(\mu_0 H_{eff})$ at high-field regime (see Fig. 3(b)), namely, the slopes and y-axis intercepts correspond to the $R_0$ and $\rho_{xy}^A$, respectively. As shown in Fig. 3(c), positive sign of $R_0$ in all temperatures reveals the dominant hole-type charge carrier. The carrier density $n_a$ can be deduced by the single band model $n_a \sim -1/|e|R_0$, which reaches $5.6 \times 10^{21}$ $cm^{-3}$ at 10 K corresponding to ~1.1 carriers per formula unit of LaMn$_2$Ge$_2$. The extracted $R_S$ ~0.2 cm$^3$/C at 300 K from $\rho_{xy}^A = R_S \mu_0 M$ is about two order magnitude larger than that of usual FM materials, such as pure Fe and Ni (*T* = 300 K) [35,36]. Furthermore, both $R_S$ and $\rho_{xy}^A$ show broad hump at 280 K [see Fig. S2], this temperature is well below $T_C$ [~0.85 $T_C$] similar to the observations in La$_{1-x}$Sr$_x$MnO$_3$ and Fe$_3$GeTe$_2$ single crystals [37,38].

To help to identify the dominant contribution of AHE, anomalous hall conductivity (AHC) $\sigma_{xy}^A$ is evaluated by the relation $\sigma_{xy}^A = |\rho_{xy}/(\rho_{xx}^2 + \rho_{xy}^2)|$. As shown in Fig. 3(d) and Fig. S2, $\sigma_{xy}^A$ at 10 K reaching up ~ 730 $\Omega^{-1}cm^{-1}$ is close to the expected value from intrinsic Berry curvature contribution $\sigma_{xy,in}^A = e^2/(ha_z) \sim 704(2)$ $\Omega^{-1}cm^{-1}$ [5,39], where *e* is the electronic charge, *h* is the Plank constant and $a_z$=c/2~5.48(1) Å is the Mn interlayer's distance. This large $\sigma_{xy}^A$ cannot be from the extrinsic side-jump mechanism because it produces $\sigma_{xy}^A$ in the order of $\frac{e^2}{ha_z}(\frac{E_{SO}}{E_F})$ ($E_{SO}$ is the energy of SOC, $E_F$ is the Fermi energy), where $\frac{E_{SO}}{E_F}$ usually has a value of $10^{-1}$-$10^{-3}$ for metallic ferromagnets [3,7]. In terms of the extrinsic skew-scattering mechanism, $\sigma_{xy}^A$ only can induce $e^2/(ha_z)$ in the ultra-clean limit with $E_{SO} \gg \hbar/\tau$ ($E_{SO}$ is the energy of SOC, $\tau$ is the scattering time), whose typical conductivity ($G_{xx}$) is ~$10^{-1}$ (S/sheet). [40] However, LaMn$_2$Ge$_2$ has the conductivity of $G_{xx}$ ~$5\times10^{-3}$ (S/sheet) at 10 K in the moderately dirty regime ($E_{SO} \sim \hbar/\tau$). Additionally, the scaling coefficient $S_H = \mu_0 R_S/\rho_{xx}^2 = \sigma_{xy}^A/M$ corresponds to the sensitivity of AHC with respect to $M$. As shown in Fig. 3(e), the derived large value of $S_H$ ~0.22(2) V$^{-1}$ at 10 K also supports that LaMn$_2$Ge$_2$ in in the moderately dirty region with relatively high conductivity in ferromagnets such as Fe$_3$Sn$_2$ ($S_H$~0.1) and PrAlGe ($S_H$~0.15 V$^{-1}$) [41,42]. In this regime, the skew-scattering contribution is much smaller than the intrinsic mechanism. Moreover, the variation of $\log \rho_{xy}^A$ versus $\log \rho_{xx}$ fitted by formula $\rho_{xy}^A \propto \rho_{xx}^\alpha$ gives the scaling exponent α=1.81(~2) (see inset of Fig. 3(e)), ruling out the skew-scattering mechanism with a relation $\rho_{xy}^A \propto \rho_{xx}$ [6]. To further check whether $\rho_{xy}^A$ has a quadratic dependence on $\rho_{xx}$, we plot $\rho_{xy}^A/\mu_0 H_{eff}$ versus $\rho_{xx}^2 M/\mu_0 H_{eff}$ at several temperatures [in Fig. 3(f)]. For clarification, the curves have been offset subsequently by 0.02 cm$^3$C$^{-1}$. The good linear scaling of the data supports that AHE is dominated by the intrinsic contribution.

Now, we turn to the Hall effect for field oriented within the *ab* plane. Fig. 4(a)-4(l) show the isothermal Hall resistivity $\rho_{zx}(\mu_0 H_{eff})$ data measured under *I // c* and *H // b*, the black lines represent the fitting curves by relation $\rho_{zx} = R_0 \mu_0 H + S_H \rho_{zz}^2 M$ with fitting parameter $R_0$ and $S_H$. At $T$ = 350 K ($T > T_C$), linear field dependence of $\rho_{zx}(\mu_0 H_{eff})$ indicates the dominant normal Hall effect (NHE). As $T$ approaches $T_C$, $\rho_{zx}(\mu_0 H_{eff})$ start to deviate from the scaling of $M(\mu_0 H_{eff})$ curves. Below $T_C$, this deviation becomes more obvious. More importantly, $\rho_{zx}(\mu_0 H_{eff})$ exhibit nontrivial field dependence with an abnormal broad peak at low-field regions ($H < H_s^a$) at which no anomaly observed in $M(\mu_0 H_{eff})$ curves. This unconventional behaviors strongly support the existence of an additional THE response besides the NHE and AHE contributions, different from the Hall response well described by the NHE and AHE contributions for *H // c*. By including THE, the total Hall resistivity $\rho_{zx}$ of LaMn$_2$Ge$_2$ can be depicted by $\rho_{zx} = \rho_{zx}^N + \rho_{zx}^A + \rho_{zx}^T = R_0 \mu_0 H + S_H \rho_{zz}^2 M + \rho_{zx}^T$, where $\rho_{zx}^N$, $\rho_{zx}^A$, and $\rho_{zx}^T$ represent the normal, anomalous and topological Hall resistivity, respectively. Since $\rho_{zx}^T$ should vanish when the full-polarized FM state is established as field above critical field ($\mu_0 H_S^T$), we can determine the coefficients $R_0$ and $S_H$ as the slope and intercept of the curve $\rho_{zx}$ vs $\rho_{zz}^2 M$ above the critical field $\mu_0 H_S^a$. Using this analysis, $\rho_{zx}^A$ (second term in above equation, $\rho_{zx}^A = S_H \rho_{zz}^2 M$) can be obtained. Afterwards, $\rho_{zx}^T$ component is separated by subtracting the normal and anomalous parts from measured $\rho_{zx}(\mu_0 H_{eff})$ curves [see Fig. S3]. The maximum amplitude of THE ($\rho_{Max}^T$) and AHE ($\rho_{Max}^A$) as function of temperature are summarized in Fig. 5(a) and 5(b). As seen, $\rho_{zx}^A$ reaches maximum at the field ($\mu_0 H_T^{Max}$) smaller than $\mu_0 H_s^a$, reveal the THE emerges during spin-flop process. Remarkably, large topological Hall resistivity $\rho_{Max}^T \sim 1.0$ $\mu\Omega$cm over broad temperature regions (190 K< $T$< 300 K) is obtained, this value is in the same order of magnitude as recent report in bulk Heusler compound $Mn_2PtSn$ (1.5 $\mu\Omega \cdot cm$) [43] and triangular-lattice $Gd_3PdSi_3$ magnet (2.6 $\mu\Omega \cdot cm$) [25] but is nearly 20 times larger than that in non-collinear antiferromagnetic $Mn_5Si_3$ films (~0.05$\mu\Omega \cdot cm$) [19]. Moreover, the window temperature of THE in $LaMn_2Ge_2$ is enhanced up to RT, making it more attractive for potential spintronic applications. To more clearly present the variation of THE, the contour plot of field-temperature (*H-T*) phase diagram using extracted THE data are constructed, shown in Fig.6 (a). Nonzero $\rho_{zx}^T$ is detected at low-field regions ($\mu_0 H < \mu_0 H_S^T$) and high temperatures. As decreased temperature, $\rho_{zx}^T$ becomes zero at low temperatures (*T*< ~70 K).

  As for the origin of THE, it is generally attributed to the spin chirality by non-coplanar spin textures [13-21]. For $LaMn_2Ge_2$, previous neutron diffraction studies reveal it forms noncollinear conical magnetic structures at zero field for $T < T_C$ [31], where the FM component is along longitudinal *c* axis (spin propagation direction) and the helical AFM component lied within the basal *ab* plane presented in Fig. 6(b)[31]. As applied *H* along the *ab* plane, the conical spin configurations on Mn sublattice will flop from *c* axis to *ab* plane, then the axis of spiral structure has a tilting toward *ab* plane and magnetic moments gain the net component in *ab* plane leading to the formation of non-coplanar spin configurations, it further evolve into possible transverse conical magnetic structure [See Fig. 6(b)]. During the spin flop, due to the presence of tilted angles between adjacent Mn layers in the projected *ab* plane [see the left side of Fig. 6(b)], nonzero chiral spin configurations will be formed, producing an emergent magnetic field ($B_{eff}$) for conductive electrons responsible for the THE. The amplitude of $B_{eff}$ depends on the configuration of spins on square lattice. Considering that topological Hall resistivity generated by $B_{eff}$ can be expressed by $\rho_{zx}^T = P R_0 B_{eff}$ [13,44], where *P* is the spin polarization of charge carriers and $R_0$ denotes the normal Hall coefficient. At 300 K, the maximum value of $B_{eff}$ can be estimated ~95 T if we take *P*~0.65 at $H_T^{Max}$ and $R_0$=0.016 $\mu\Omega \cdot cm/T$ [see Fig. S1]. At $H_T^{Max}$, THE reaches maximum. Further increased field above $\mu_0 H_s^T$, a full polarized FM state is established, leads to the vanishing of $B_{eff}$ and THE. Thus, $\mu_0 H_S^T$ determines the phase boundary of magnetic transition from the noncoplanar spin configurations to polarized FM states in the *H-T* phase diagram. Additionally, for the origin of THE, we would like to highlight the thermal-driven chiral spin fluctuation mechanism but not static non-coplanar spin configurations as report in triangular-lattice Gd$_3$PdSi$_3$ [25] and Fe$_3$GeTe$_2$ magnets [38], since the THE in LaMn$_2$Ge$_2$ is only observed at high temperatures instead of low temperatures. Similar nontrivial skyrmion-lattice textures stabilized by thermal fluctuation in MnSi [45] and chiral spin fluctuations in SrRuO$_3$ films [46] have been proposed for the THE. Here, we propose that transverse conical spin fluctuations may be responsible for the THE in LaMn$_2$Ge$_2$ at elevated temperatures.

In comparison with the low temperature THE observed in majority of frustrated magnets, the realization of RT THE is critical to design novel spintronic devices based on THE. Here, the observed RT THE in LaMn$_2$Ge$_2$ is quite large when compared to the limited materials hosting THE near RT, such as bulk Mn$_3$Sn [47], MnPdGa [48], Mn$_2$NiGa [49] and Fe$_3$Sn$_2$ [28], films of Mn$_2$PtSn [50], FeGe [51], and MnNiGa [52] [shown in Fig. 7]. In LaMn$_2$Ge$_2$, THE component dominates the Hall effect at low fields while AHE reaches maximum at high-field region. More importantly, THE has a much larger value than AHE, as example, the calculated maximum topological Hall conductivity (THC) $\sigma_{Max}^T \sim 45 \ \Omega^{-1} cm^{-1}$ is four times the value of AHC $\sigma_{Max}^A \sim 12 \ \Omega^{-1} cm^{-1}$ at 300 K [see Fig. S3]. This is also different from the Hall effect observed in most Heusler compounds hosting Skyrimion lattice where the AHE is usually larger than THE [48-51]. Thus, it is a promising materials for technological applications in spintronic device based on THE.

## IV. CONCLUSIONS

In summary, we reported the strong anisotropic Hall effect in a noncollinear ferromagnetic LaMn$_2$Ge$_2$ with $T_C \sim 325$ K. As *H // c* and *I // ab*, it exhibits the conventional AHE dominated by the intrinsic Berry curvature mechanism. While, for *H // ab* and *I // c,* large topological Hall resistivity in a broad field and temperature (190 K<*T*<310 K) window can be observed, which arise from the formation of noncoplanar spin configuration with finite chirality stabilized by thermal fluctuations. The realization of large THE near RT in LaMn$_2$Ge$_2$ entails it as candidate material for practical applications in the THE-based spintronic devices.

### ACKNOWLEDGMENTS

This work is supported by the National Natural Science Foundation of China (Grant No. 11874158), and Fundamental Research Funds for the Central Universities (Grant No.

2018KFYYXJJ038 and 2019KFYXKJC008). We would like to thank the staff of the analysis center of Huazhong University of Science and Technology for their assistance in structural characterization and analysis.

[1] E. H. Hall, Am. J. Math. **2**, 287 (1879).
[2] E.H. Hall, Philos. Mag. **12**,157 (1881).
[3] N. Nagaosa, J. Sinova, S. Onoda, A. H. MacDonald, and N. P. Ong, Rev. Mod. Phys. **82**, 1539 (2010).
[4] S. Onoda, N. Sugimoto, and N. Nagaosa, Phys. Rev. Lett. **97**, 126602(2006).
[5] D. Xiao, M.C. Chang, and Q. Niu, Rev. Mod. Phys. **82**,1959 (2010).
[6] J. Smit, Physica **24**, 39 (1958).
[7] L. Berger, Phys. Rev. B **2**, 4559 (1970).
[8] R. Karplus, J. M. Luttinger, Phys. Rev. **95**,1154(1954).
[9] G. Sundaram and Q. Niu, Phys. Rev. B **59**, 14915 (1999).
[10] T. Jungwirth, Q. Niu and A. H. MacDonald, Phys.Rev. Lett. **88**, 207208(2002).
[11] Y. Taguchi, Y. Oohara, H. Yoshizawa, N. Nagaosa, and Y. Tokura, Science **291**, 2573 (2001).
[12] Y. Machida, S. Nakatsuji, Y. Maeno, T. Tayama, T. Sakakibara, and S. Onoda, Phys. Rev. Lett. **98**, 057203 (2007).
[13] A. Neubauer, C. Pfeiderer, B. Binz, A. Rosch, R. Ritz, P.G. Niklowitz, and P. Böni, Phys. Rev. Lett. 102, 186602 (2009).
[14] T. Schulz, R.Ritz, A. Bauer, M. Halder, M. Wagner, C. Franz, C. Pfeiderer, K. Everschor, M. Garst, and A. Rosch, Nat. Phys. **8**, 301 (2012).
[15] N. Kanazawa, Y. Onose, T. Arima, D. Okuyama, K. Ohoyama, S. Wakimoto, K. Kakurai, S. Ishiwata, Y. Tokura, Phys. Rev. Lett. **106**, 156603 (2011).
[16] L. M. Wang, Phys. Rev.Lett. **96**, 077203 (2006).
[17] L. Vistoli, W. Wang, A. Sander, Q. Zhu, B. Casals, R. Cichelero, A. Barthélémy, S. Fusil,G. Herranz, S. Valencia, R. Abrudan, E. Weschke, K. Nakazawa, H. Kohno, J. Santamaria, W. Wu, V. Garcia, and M. Bibes, Nat. Phys. **15**, 67 (2019).
[18] Y. Shiomi, M. Mochizuki, Y. Kaneko, and Y. Tokura, Phys.Rev. Lett. **108**, 056601(2012).
[19] C. Sűrgers, G. Fischer, P. Winkel, and H. V. Löhneysen, Nat. Comm. **5**, 3400 (2014)
[20] T. Suzuki, R. Chisnell, A. Devarakonda, Y.T. Liu, W. Feng, D. Xiao, J.W. Lynn, J.G. Checkelsky, Nat. Phys. **12**, 1121 (2016).
[21] H. Li, B. Ding, J. Chen, Z. Li, E. Liu, X. Xi, G. Wu, W. Wang, Appl. Phys. Lett. **116**, 182405 (2020).
[22] Q. Shao, Y. Liu, G. Yu, S. Kim, X. Che, C. Tang, Q. He, Y. Tserkovnyak, J. Shi, and K. Wang, Nat. Electron. **2**, 182 (2019).
[23] J. Matsuno, N. Ogawa, K. Yasuda, F. Kagawa, W. Koshibae, N. Nagaosa, Y. Tokura, and M. Kawasaki, Sci. Adv. **2**, e1600304 (2016).
[24] H. Takatsu, S. Yonezawa, S. Fujimoto, and Y. Maeno, Phys. Rev. Lett. **105**, 137201 (2010).
[25] T. Kurumaji, T. Nakajima, M. Hirschberger, A. Kikkawa, Y. Yamasaki, H. Sagayama, H.

Nakao, Y. Taguchi, T. Arima, Y. Tokura, Science **365**, 914 (2019).

[26] X. Li, L. Xu, L. Ding, J. Wang, M. Shen, X. Lu, Z. Zhu, and K. Behnia, Phys. Rev. Lett. **119**, 056601 (2017).

[27] P. K. Rout, P.V.P. Madduri, S.K. Manna, A. K. Nayakx, Phys. Rev. B **99**, 094430 (2019).

[28] H. Li, B. Ding, J. Chen, Z. Li, Z. Hou, E. Liu, H. Zhang, X. Xi, G. Wu, W. Wang, Appl. Phys. Lett. 114, 192408 (2019).

[29] O. Masaru, T.Gen, and N. Naoto, J. Phys. Soc. Jpn. **73**, 2624 (2004).

[30] Z. Ban and M. Sikirica, Acta Crystallogr. **18**, 594 (1965).

[31] G. Venturini, R. Welter, E. Ressouche, and B. Malaman, J. Alloy & Compd, **210**, 213 (1994).

[32] A. Aharoni, J. Appl. Phys. **83**, 3432 (1998).

[33] T. Shigeoka, N. Iwata, H. Fujii, and T. Okamoto, J. Mag & Mag Mater. **53**, 83 (1985).

[34] See Supplemental Material for the temperature dependence of magnetoresistance, normal hall coefficient and anomalous Hall resistiviy with field along different directions, and the isothermal anomalous Hall resistivity and topological Hall resistivity at different temperatures.

[35] N. V. Volkenshtein, G. V. Fedorov, Sov. Phys. JETP **11**, 48 (1960).

[36] S. N. Kaul. Phys. Rev. B **20**, 5122 (1979).

[37] Y. Onose and Y. Tokura, Phys. Rev. B **73**, 174421(2006).

[38] Y. Wang, C. Xian, J. Wang, B. Liu, L. Ling, L. Zhang, L. Cao, Z. Qu, and Y. Xiong, Phys. Rev B **96**, 134428 (2017).

[39] K. Manna, Y. Sun, L. Muechler, J. Kübler, and C. Felser, Nat. Rev. Mater, **3**, 244 (2018).

[40] S. Onoga, N. Sugimoto, and N. Nagaosa, Phys. Rev. B **77**, 165103(2008)

[41] Q. Wang, S. Sun, X. Zhang, F. Pang, and H. Lei, Phys. Rev. B **94**, 075135(2016)

[42] B. Meng, H. Wu, Y. Qiu, C. Wang, Y. Liu, Z. Xia, S. Yuan, H. Chang, and Z. Tian, APL Mater. **7**, 051110 (2019).

[43] Z. Liu, A. Burigu, Y. Zhang, H. Jafri, X. Ma, E. Liu, W. Wang, and G. Wu, Scrip. Mater. 143, 122 (2018).

[44] R. Ritz, M. Halder, C. Franz, A. Bauer, M. Wagner, R. Bamler, A. Rosch, and C. Pfleiderer, Phys.Rev. B **87**, 134424 (2013).

[45] S. Mühlbauer, B. Binz, F. Jonietz, C. Pfleiderer, A. Rosch, A. Neubauer, R. Georgii, and P. Böni, Science **323**, 915 (2009).

[46] W. Wang, M.W. Daniels, Z. Liao, Y. Zhao, J. Wang, G. Koster, G. Rijnders, C. Chang, D. Xiao, and W. Wu, Nat. Mater. **18**,1054 (2019).

[47] X. Li, C. Collignon, L. Xu, H. Zuo, A. Cavanna, U. Gennser, D. Mailly, B. Fauqué, L. Balents, Z. Zhu, and K. Behnia, Nat. Comm. **10**, 3022 (2019).

[48] X. Xiao, L. Peng, X. Zhao, Y. Zhang, Y. Dai, J. Guo, M. Tong, J. Li, B. Li, W. Liu, J. Cai, B. Shen, and Z. Zhang, Appl. Phys. Lett. **114**, 142404 (2019).

[49] S. Sen, C. Singh, P. K. Mukharjee, R. Nath, and A. K. Nayak, Phys. Rev. B **99**, 134404 (2019).

[50] Y. Li, B. Ding, X. Wang, H. Zhang, W. Wang, and Z. Liu, Appl. Phys. Lett. **113**, 062406 (2018).

[51] J.C. Gallagher, K. Y. Meng, J. T. Brangham, H.L. Wang, B. D. Esser, D. W. McComb,

and F. Y. Yang, Phys. Rev. Lett. **118**, 027201(2017).

[52] B. Ding, Y. Li, G. Xu, Y. Wang, Z. Hou, E. Liu, Z. Liu, G. Wu, and W. Wang, Appl. Phys. Lett. **110**, 092404 (2017).

**Figure Captions**

FIG. 1 (a) The crystal structure of LaMn$_2$Ge$_2$. Purple, green and blue balls represent the Mn, Ge and La atoms, respectively. (b) Top view of square layer of Mn atoms, (c) Temperature dependence of susceptibility χ(T) with ZFC and FC modes under $\mu_0H$ = 0.1 T for *H* // *c* and *H* // *a* axis, respectively. (d) Temperature dependence of in-plane ($\rho_{xx}$) and out-of-plane ($\rho_{zz}$) longitudinal resistivity for *I* // *a* and *c* axis, respectively.

FIG. 2 The isothermal magnetization $M(\mu_0H_{eff})$ curves at different temperatures for (a) *H* // c axis and (b) *H* // *a* axis, respectively.

FIG. 3 Field dependence of (a) magnetoresistance and (b) Hall resistivity $\rho_{xy}$ at selected temperatures under *H*// c axis and *I* // *a* axis, the inset shows the schematic setup of Hall resistivity measurements. (c) Temperature dependence of ordinary Hall coefficient $R_0$ and anomalous Hall coefficient $R_S$. (d) Temperature dependence of $\sigma_{xy}^A$, inset shows the plot of $\log\rho_{xy}$ versus $\log\rho_{xx}$. (e) Temperature dependence of scaling coefficient $S_H$. (f) The scaling behavior of $\rho_{xy}/\mu_0H_{eff}$ versus $\rho_{xx}^2 M/\mu_0H_{eff}$ at indicated temperatures with subsequent offset of 0.02 cm$^3$C$^{-1}$ for clarity, solid lines represent linear fits of the data.

FIG. 4 (a-l) Isothermal Hall resistivity $\rho_{zx}$ at various temperatures under *H* // *a* axis and *I* // *c* axis, respectively. The black solid lines are the fitting curves including normal and anomalous Hall contributions.

FIG. 5 (a) Temperature dependence of maximum amplitude of THE ($\rho_{\text{Max}}^{T}$) and corresponding field position ($\mu_0H_C^{\text{Max}}$), (b) Temperature dependence of maximum values of AHE ($\rho_{\text{Max}}^{A}$) and saturated field ($\mu_0H_s^{ab}$),

FIG. 6 (a) Contour plots of topological Hall resistivity versus temperature and field for LaMn$_2$Ge$_2$ under *H* // *a* axis. The transition between noncoplanar spin structure and polarized FM order is marked by orange cycles ($\mu_0H_s^{ab}$), (b) The Mn moment projects for adjacent layers within *ab* plane (left side), *ϕ* is the angle between two magnetic moments at "1" and "1'" sites in the *ab* plane. The schematic spin configurations of Mn moments at $\mu_0H$ = 0 T, 0 < $\mu_0H$ < $\mu_0H_s^{ab}$ and $\mu_0H$ > $\mu_0H_s^{ab}$ for *H* // *a* axis (right side).

FIG. 7 Comparisons of $\rho_{\text{Max}}^{T}$ for various materials exhibiting RT THE related to the non-coplanar spin textures. The data are taken from the open literatures, including Refs. 28

and 47-52. The vertical dashed line represents the position of 300 K.

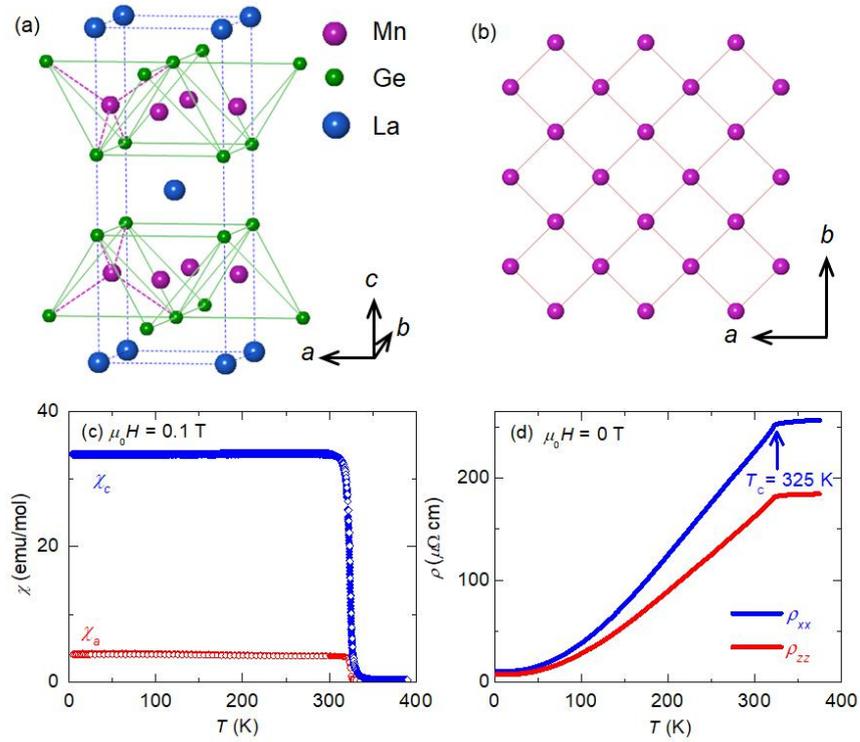

FIG. 1

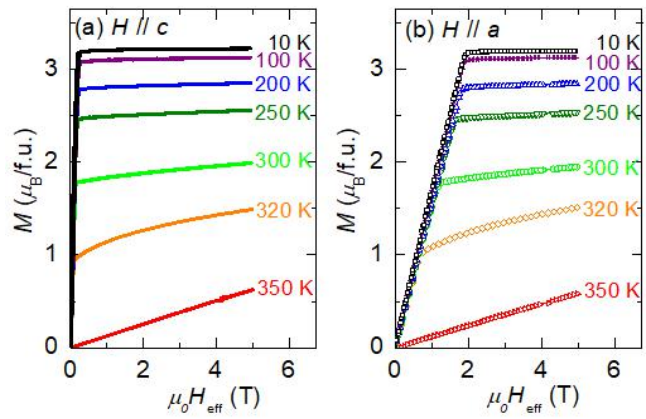

FIG. 2

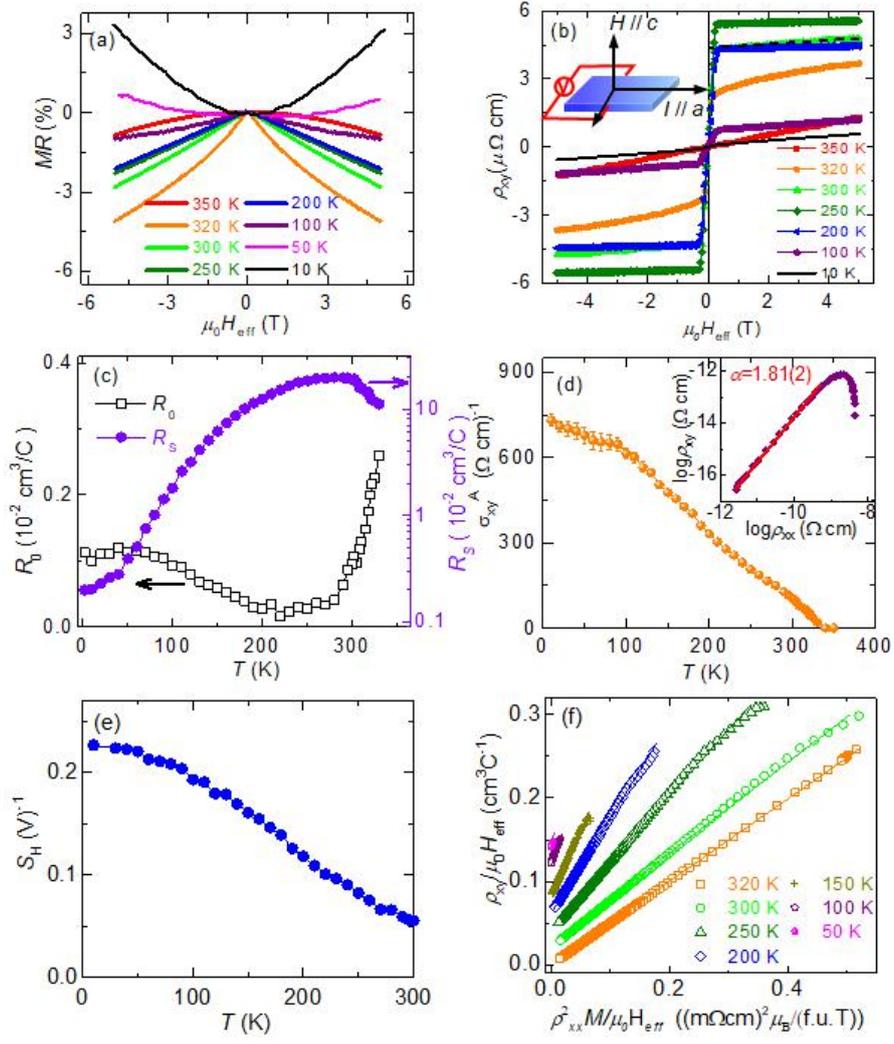

FIG. 3

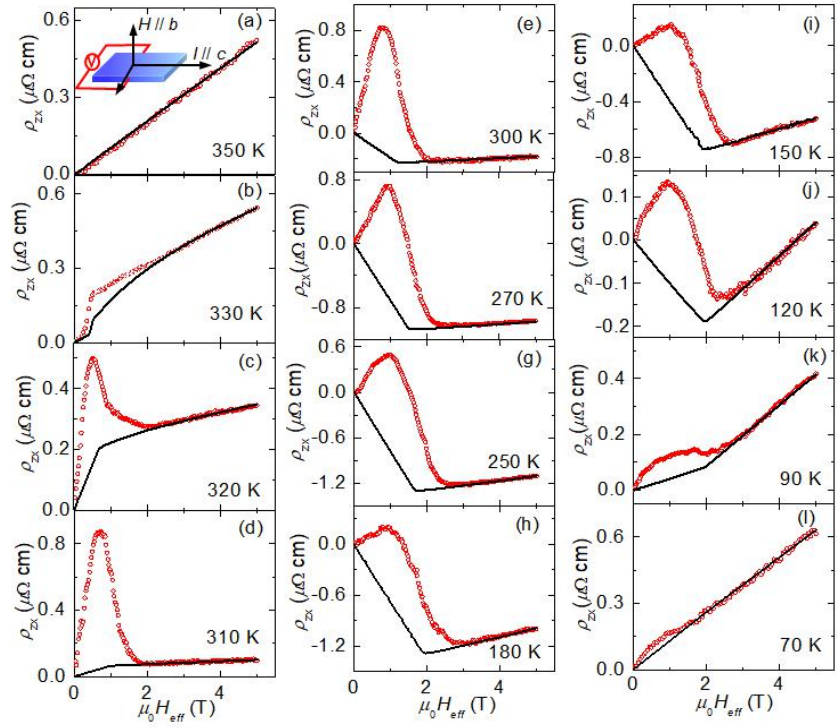

FIG. 4

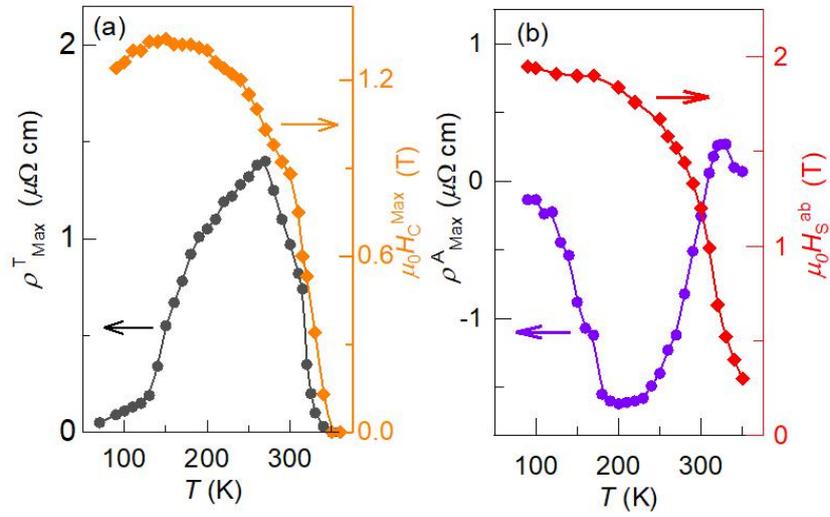

FIG.5

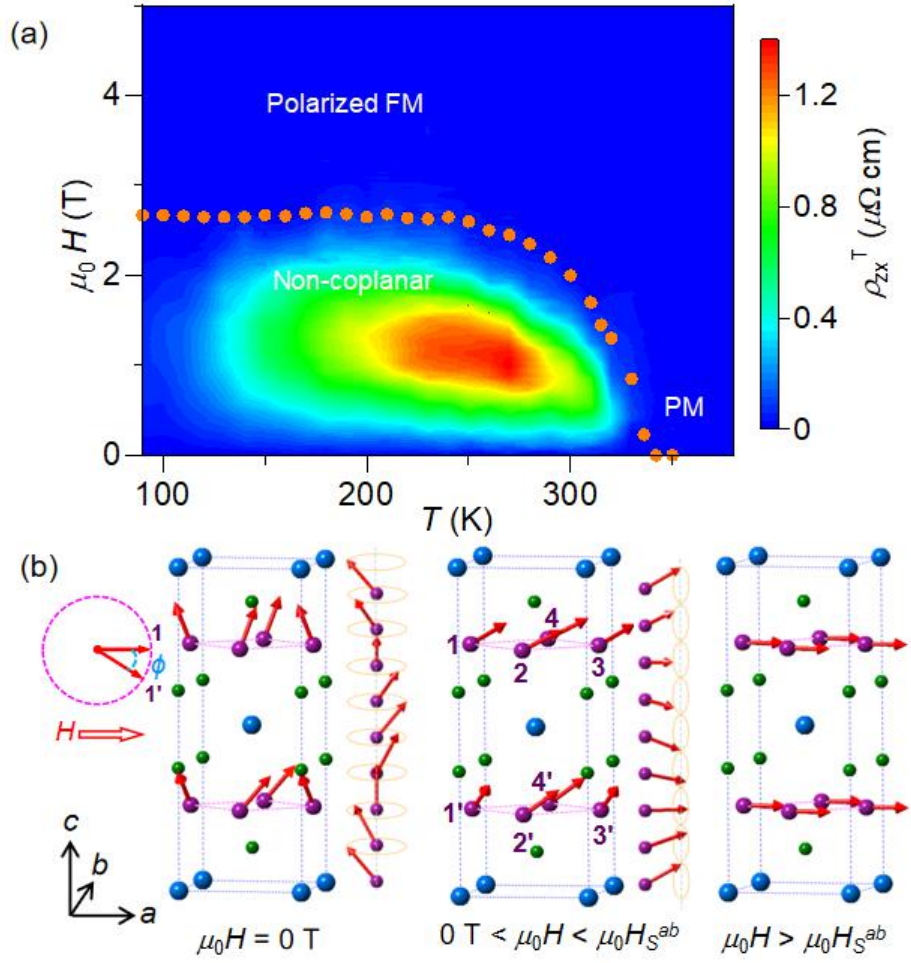

FIG.6

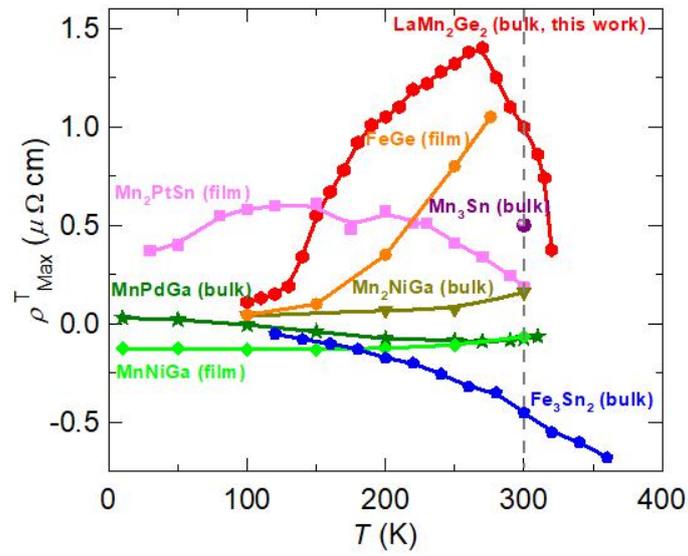

FIG. 7